\documentclass[11pt,a4paper]{article} 
\usepackage{jcappub}
\usepackage{bm}
\usepackage{times} 
\usepackage{amsmath} 
\usepackage{color}

\newcommand{\up}[1]{{\rm #1}}
\newcommand{\bdi}[1]{\hbox{\boldmath{$#1$}}}
\newcommand{\beeq}{\begin{equation}}
\newcommand{\eneq}{\end{equation}}
\newcommand{\bear}{\begin{eqnarray}}
\newcommand{\enar}{\end{eqnarray}}
\newcommand{\AVE}[1]{\langle#1\rangle}
\newcommand{\rbar}{\bar r}   
\newcommand{\ra}{\rightarrow}       
\newcommand{\ax}{\alpha_{\chi}}  
\newcommand{\px}{\varphi_{\chi}} 
\newcommand{\zz}{z}              
\newcommand{\ttt}{\theta}        
\newcommand{\pp}{\phi}           
     
\newcommand{\drr}{\delta r}    
\newcommand{\dtt}{\delta\ttt}  
\newcommand{\dpp}{\delta\pp}   
\newcommand{\dT}{\delta\eta}   
\newcommand{\dz}{\delta z}     
\newcommand{\HH}{\mathcal{H}}  
\newcommand{\DT}{\Delta\eta}   
\newcommand{\DX}{\Delta x}     
\newcommand{\dL}{\mathcal{D}_L}  
\newcommand{\ddL}{\delta\mathcal{D}_L} 
\newcommand{\LL}{\mathcal{L}}
\newcommand{\OO}{\mathcal{O}}
\newcommand{\Sb}{\mathbb{S}}

\begin{document}

\begin{titlepage}

\setcounter{page}{1} \baselineskip=15.5pt \thispagestyle{empty}
\pagenumbering{roman}

\bigskip

\vspace{1cm}
\begin{center}
{\fontsize{20}{28}\selectfont \bfseries Gauge-Transformation Properties of
Cosmological Observables and its Application to the Light-Cone Average}
\end{center}

\vspace{0.2cm}

\begin{center}
{\fontsize{13}{30}\selectfont Jaiyul Yoo,$^{a,b}$ and Ruth Durrer $^c$}
\end{center}

\begin{center}
\vskip 8pt
\textsl{$^a$ Center for Theoretical Astrophysics and Cosmology,
Institute for Computational Science}\\
\textsl{University of Z\"urich, Winterthurerstrasse 190,
CH-8057, Z\"urich, Switzerland}

\vskip 7pt

\textsl{$^b$Physics Institute, University of Z\"urich,
Winterthurerstrasse 190, CH-8057, Z\"urich, Switzerland}

\vskip 7pt

\textsl{$^c$D\'epartement de Physique Th{\'e}orique \& Center 
for Astroparticle Physics, Universit\'e de Gen\`eve\\
Quai E. Ansermet 24, CH-1211 Gen\`eve 4, Switzerland} \vspace{0.3cm}\\
\today 
\end{center}

\note{jyoo@physik.uzh.ch, ~~~ruth.durrer@unige.ch}

\vspace{1.2cm}
\hrule \vspace{0.3cm}
\noindent {\sffamily \bfseries Abstract} \\[0.1cm]
Theoretical descriptions of observable quantities in cosmological 
perturbation theory should be independent of coordinate systems. This statement
is often referred to as  gauge-invariance of observable quantities,
and the sanity of their theoretical description is verified by checking
its gauge-invariance. We argue that cosmological observables are 
invariant scalars under  diffeomorphisms and their
theoretical description is gauge-invariant, {\it only}
at linear order in perturbations. Beyond linear order,
they are usually not gauge-invariant, and 
we provide the general law
for the gauge-transformation that the perturbation part  of an observable
 does obey. We apply this finding
to derive the second-order expression for the observational light-cone average 
in cosmology and
demonstrate that our expression is indeed invariant under diffeomorphisms.
\vskip 10pt
\hrule

\vspace{0.6cm}
\end{titlepage}

\noindent\hrulefill

\tableofcontents

\noindent\hrulefill

\pagenumbering{arabic}

\section{Introduction}
In the past few decades, cosmological observations firmly established
that,  on sufficiently large scales,  our Universe is well approximated by a homogeneous and isotropic expanding
Universe described by a Friedmann-Lema\^\i tre-Robertson-Walker (FLRW) spacetime with small perturbations.
 In particular,  linear-order cosmological
perturbation theory has been extensively used to model
various cosmological
observables beyond the background predictions such as galaxy clustering
on large scales
and anisotropies of the cosmic microwave background, which play an essential role
in establishing the standard cosmological model. In the coming years,
cosmology research is poised to take a further leap and enter into
an era of truly precision cosmology with numerous
large-scale surveys in the offing. Surveys such as 
the Dark Energy Spectroscopic Instrument,\footnote{http://desi.lbl.gov/}
Euclid,\footnote{http://sci.esa.int/euclid/}
WFIRST,\footnote{https://wfirst.gsfc.nasa.gov/}
and the Large Synoptic Survey Telescope\footnote{https://www.lsst.org/}
will deliver
unprecedented amounts of data with equally unprecedented precision. These will allow
measurements of
higher-order statistics, which require higher-order perturbation theory
for their predictions, and which are expected to provide additional information about
the nature of gravity and the initial conditions of the early Universe.

Together with this rapid development in observations and experiments, 
a large amount of work has been devoted to developing a solid theoretical framework
for predictions of cosmological observables. In particular, several
higher-order relativistic calculations of cosmological observables
have been performed in recent years,
see, e.g., \cite{BEGAET12a,BEGAET13,UMCLMA14a,BEDUET14}
for the luminosity distance
and \cite{YOZA14,YOO14b,BEMACL14,BEMACL14b,DIDUET14,DiDio:2015bua}
for galaxy clustering. 
As a consequence of diffeomorphism invariance of the underlying theory, 
these general relativistic calculations can be performed in any coordinate 
system at any order. 
One subtlety in cosmology exists ---
due to the symmetry in the FLRW metric, it is convenient to
split physical quantities into a background contribution and a perturbation. 
This split is however to some degree arbitrary. 
Only the full, perturbed spacetime is physical and not the background neither 
its perturbation. A gauge transformation is actually a (small) change 
in this split.
To first order, this is equivalent to a coordinate transformation
 $x^a\mapsto x^a +\epsilon~ \xi^a$ under which any arbitrary variable 
${ O} = \bar{O}  + \epsilon\delta{ O} $ changes by the Lie derivative,
$$ O \mapsto \bar{O}  + \epsilon(\delta{O} - {\cal L}_\xi\bar{O}) \,.$$
This is a first order gauge transformation~\cite{Hawking:1966qi,BARDE80,KOSA84} and it can be easily generalized to higher orders, see~\cite{BRMAET97,BRSO99} and Section~\ref{ss:obs-gi}.

Given  diffeomorphism invariance in general relativity, perturbations
change (or gauge-transform), while  cosmological observables
such as galaxy clustering or the cosmic microwave background anisotropies
are  independent of our coordinate choice. This statement is often
phrased as the gauge-invariance of the cosmological observables, and it might be {\it expected} to apply to perturbation calculations at {\it all orders.}
The fact that, at first order,  cosmological observables can be expressed in a gauge-invariant way 
puts  tight constraints on their expressions in terms of perturbation variables, 
providing a useful way to verify the validity of complicated relativistic
calculations. 
This point indeed has been adopted in the past to successfully develop the
relativistic formalism for galaxy clustering and gravitational lensing
(see, e.g., 
\cite{YOFIZA09,YOO10,YOO14a,BODU11,CHLE11,JESCHI12,BRCRET12,
SCJE12a,BEBOET12}), 
in which the theoretical expressions for these cosmological
observables are expressed in terms of general metric perturbations
and are shown to remain invariant under gauge transformations.
The caveat is that these gauge-invariant calculations in the past are
performed only at the {\it linear-order}.  As we shall argue here, beyond linear order gauge-invariance is no longer a useful criterion.

Calculations of cosmological observables are much more complicated
beyond the linear order in perturbations, and  the gauge-transformation properties of
second-order (or higher-order) expressions 
are expected to play a critical role in both
verifying the complicated calculations and 
reaching a consensus among different practitioners in the field.
However, as we clarify in this paper, cosmological observables are
diffeomorphism invariant scalars, {\it not} gauge-invariant.
 The statement of gauge-invariance of cosmological observables is
{\it incorrect} beyond linear order and the
theoretical expressions for cosmological observables {\it do}
gauge transform, starting at  second order in perturbations.

In Section~\ref{sec:cosobs}, 
we demonstrate that cosmological observables can be expressed as
scalars under diffeomorphisms.
Consequently, {\it diffeomorphism invariance} of cosmological observables is the condition to be used to test theoretical consistency rather than gauge-invariance.
Starting from this condition, we show that the theoretical expressions are 
in fact  gauge-invariant
at  linear order in perturbations, and we provide
the transformation properties  beyond linear order.
With a few assumptions,
the same conclusion was reached \cite{BRSO99} (see also \cite{BRMAET97}),
in which the gauge-transformation properties of  cosmological observables are derived order by order in perturbations. We comment on this early work
in Section~\ref{sec:discuss}.

In this paper we mainly demonstrate a new proposal on  how to check the 
highly involved second order calculations, where gauge-invariance 
does not hold, in the future. We apply it to some simple examples 
which  are not new, they are rather an illustration of our method.
As a proof of concept, we apply our findings to derive the second-order
expressions of the observational light-cone average of the luminosity 
distance~$\dL$ in Section~\ref{sec:obsavg}. 
The light-cone average of the luminosity distance was considered 
 previously in Refs.~\cite{BOCLET15b,FLCLMA16,KAPE16}, where  some
relativistic corrections are neglected in these works and hence the diffeomorphism
invariance of the light-cone average is broken.
We show how diffeomorphism invariance can be used to derive the
correct expression of the light-cone average at  second order in 
perturbations. 
Our expression is formulated such that it can easily
be generalized to any other observable on the light cone.

The luminosity distance~$\dL$ is constructed
by measuring the flux of  standard candles like supernovae at the
observed redshift~$z$ and the angular direction~$\bdi{n}$. These measurements
are often averaged over directions~$\bdi{n}$ at the same redshift~$z$
to obtain the luminosity-redshift relation as
\beeq\label{e:av1}
\left\langle\dL(z,\bdi{n})\right\rangle_\up{obs}
\equiv{1\over N_g}\sum_{i=1}^{N_g}\dL(z,\bdi{n}_i)~,
\eneq
where $N_g$ is the total number of host galaxies with 
the luminosity distance measurements in the redshift bin $[z,~z+dz]$.
Since this observational light-cone
average is performed over the angle~$\bdi{n}$, it is often
referred to as the observational angular average, which is defined 
with the subscript~$\Omega$ in our notation as
\beeq\label{e:av2}
\AVE{\dL(z,\bdi{n})}_\Omega\equiv{1\over \Omega}\int d^2\bdi{n}~\dL(z,\bdi{n})~,
\qquad\qquad \Omega=\int d^2\bdi{n}~,
\eneq
where $\Omega$ is the total angular area 
the observational angular average is performed over. However, it is well
known \cite{BOCLET15b,FLCLMA16}
that beyond linear perturbation theory the two averages \eqref{e:av1} and \eqref{e:av2} are related, but not equivalent even in the limit 
$N_g\rightarrow\infty$.
Splitting each measurement of~$\dL(z,\bdi{n}_i)$ into a background~$\bar D_L(z)$
and a (dimensionless)
perturbation~$\ddL$ around it, we define the perturbation to
the observational light-cone average as
\beeq
\left\langle\ddL\right\rangle_\up{obs}
\equiv{\left\langle\dL(z,\bdi{n})\right\rangle_\up{obs}
-\bar D_L(z)\over\bar D_L(z)}=
{1\over N_g}\sum_{i=1}^{N_g}\ddL(z,\bdi{n}_i)~.
\eneq
The task of theorists is to relate this observational light-cone
average
to the angular average and eventually to an
 ensemble average, which can be
 computed  within cosmological perturbation
theory at the requested order. In Section~\ref{sec:obsavg} we compute the
observational light-cone average to the second order in perturbations
and show that it satisfies
the transformation property required to make the observational light-cone
average invariant under a general coordinate transformation.

In Section~\ref{sec:discuss} we discuss the implications of our findings,
and some detailed equations are presented in Appendix~\ref{app:metric}.
We use $a,b,c\in\{0,1,2,3\}$ to represent the spacetime indices and
  $i,j,k\in\{1,2,3\}$ to represent the spatial indices.

\section{Observable quantities in cosmology and their gauge-transformation}
\label{sec:cosobs}
We show that cosmological observables are expressed in terms of
scalars, invariant under coordinate 
transformations. With this condition for diffeomorphism invariance, 
we demonstrate
that  perturbations of cosmological observables are gauge-invariant
at  linear order in perturbations but they {\it do} gauge-transform
beyond the linear order.

\subsection{Observable quantities as scalars under a diffeomorphism}
The diffeomorphism symmetry in general relativity provides us with a 
freedom to choose a  global coordinate system.
Regardless of how we set up the  coordinate system, 
observable quantities
in cosmology should have  identical values in any of those coordinates.
This requirement implies that  scalar $S$, vector $V^a$,
or tensor $T_{ab}$ quantities describing cosmological observables
 transform as
\beeq
\label{eq:trans}
\tilde S(\tilde x)=S(x)~,\qquad \tilde V^a(\tilde x)={\partial \tilde x^a
\over\partial x^b}~V^b(x)~,\qquad \tilde T_{ab}(\tilde x)=
{\partial  x^c \over\partial\tilde x^a}{\partial  x^d \over\partial \tilde x^b}
~T_{cd}(x)~, \qquad \cdots~,
\eneq
for a given coordinate transformation (or  diffeomorphism)
\beeq
\label{eq:ct}
\tilde x^a=x^a+\xi^a~,\qquad\qquad \xi^a=(T,~L^i)~,
\eneq
describing the same physical (spacetime) point, 
while the coordinate values~$x^a$ and~$\tilde x^a$
are different in two coordinate systems.\footnote{We defined the
coordinate transformation in a non-perturbative way: $\xi^a=\epsilon\xi^{a(1)}+
\epsilon^2\xi^{a(2)}+\cdots$. There exist other definitions for the transformation
in literature using the Lie derivative.}
We suppress the coordinate indices
when used as an argument of functions.
However, these quantities describing cosmological observables
are not directly observable ---
they are in fact measured by an observer in her rest frame,
in which the four velocity $u^a(=[e_t]^a$) 
of the observer sets the time direction,
a spatial triad $[e_i]^a$ orthogonal to the time direction
defines the spatial directions, to that the local metric tensor is Minkowski,
$\eta_{IJ}=g_{ab}[e_I]^a[e_J]^b$ ($I,J=t,x,y,z$).

As an example, let us consider observing
photons emitted from a distant source and the light propagation described
by a photon wave vector~$k^a$. This observation is performed in the
observer's rest frame, and the photon wave vector is described by
two observable quantities, the photon frequency~$\nu$ ($\omega=2\pi\nu$)
and the photon propagation direction~$\bdi{n}$ as
\beeq
k^I_o=\eta^{IJ}[e_J]^bk_b
=\left(\omega~,\bdi{k}\right)=\omega~(1~,~\bdi{n})~,\qquad\qquad
\bdi{n}\equiv\bdi{k}/|\bdi{k}|~, \quad k_b = \vartheta_{,b}
\eneq
where we use the subscript~$o$ to represent  quantities
in the local rest frame of the observer and the photon wave vector is the gradient of
the phase~$\vartheta$  of the electromagnetic wave describing our light ray~\cite{SCEHFA92}. 
The photon frequency measured by the observer is a scalar under diffeomorphisms,
\beeq
-\omega=u\cdot k=\eta_{IJ}u^I_ok^J_o=g_{ab}u^ak^b~,
\eneq
and the observed direction of the light propagation is also a scalar
under diffeomorphisms
\beeq
\bdi{n}\equiv {1\over\omega} \bdi{ e}_ik_o^i \,, \qquad n^i={1\over\omega} [e_i]_ak^a~,
\eneq
where we use $\bdi{ e}_i$ to represent unit orthonormal vectors in the 
observer's rest frame ($i=x,y,z$).
The observed redshift, as another example, is simply the ratio of the photon 
frequencies (or wavelengths) in the source rest frame and the observer rest
frame:
\beeq
\label{eq:obsdef}
1+z={(u\cdot k)_s\over(u\cdot k)_o}~,
\eneq
and it is obviously a scalar under  diffeomorphisms.

All vector and tensor quantities describing cosmological observables
are measured in the observer's rest frame with respect to  local tetrads,
and hence  are indeed expressed in terms of scalars
under diffeomorphisms. The observations are
made in the unique rest frame of the observer (unique up to a global rotation of the triad $(\bdi{e}_i)$) and they are not affected
by  diffeomorphisms.  The  rest-frame of the local observer
is fully determined up to a simple spatial rotation (of the spatial triad).
This residual symmetry is internal in the sense that it is independent of the
FRW coordinate transformation. In this sense, observable quantities
are {\it diffeomorphism-invariant scalars}.

\subsection{Cosmological observables and gauge-invariance}\label{ss:obs-gi}
In cosmology, a quantity~$O$ including cosmological observables is split
into the background $\bar O$ and the perturbation $\delta O$
around it, and due to the
spatial symmetry in a homogeneous  and isotropic universe, the background quantities
depend only on time~$t$:\footnote{From now on, we absorb the small parameter $\epsilon$ in the corresponding variable and indicate the order simply by a superscript $^{(n)}$.}
\beeq
O(x)\equiv\bar O(t)(1+\delta O)~,\qquad\qquad \delta O=\delta O(t,x^i)
=\delta O^{(1)}+\delta O^{(2)}+\cdots~.
\eneq
General covariance of general relativity allows any coordinate
system to be used to describe physical phenomena, and a coordinate
transformation in Eq.~\eqref{eq:ct} necessarily involves changes in these
quantities according to Eq.~\eqref{eq:trans}. A gauge transformation, on the
other hand, is a change in the correspondence of our physical universe
to a homogeneous and isotropic background introduced to define the perturbations $\delta O$, 
while the background coordinate~$x^a$ and hence the background quantities 
$\bar O(t)$ are fixed \cite{BARDE80,KOSA86}.

Let us consider the transformation of the perturbation part of cosmological observables under the diffeomorphism given in Eq.~\eqref{eq:ct}. Since
the same physical point has two different values for its time coordinate,
the correspondence to the background in a homogeneous universe
differs in the two coordinates by
\beeq
\tilde \eta=\eta+T~,\qquad\qquad
\bar O(\tilde \eta)=\bar O(\eta)\left(1+\frac1{\bar O}{d\bar O\over d\eta}~T
+\frac1{2\bar O}{d^2\bar O\over d\eta^2}~T^2+\cdots\right)~,
\eneq
where we use conformal coordinate time~$\eta$ instead of the cosmic proper
time coordinate~$t$ and we expand $\bar O(\tilde \eta)$ in a Taylor series. Observable quantities are invariant
under  diffeomorphisms, $\tilde O(\tilde x)=O(x)$. This condition implies that
the perturbation part~$\delta O$ transforms as
\beeq
\label{eq:general}
\widetilde{\delta O}(\tilde x)=\left[1+\delta O(x)\right]
\left(1+\frac1{\bar O}{d\bar O\over d\eta}~T
+\frac1{2\bar O}{d^2\bar O\over d\eta^2}~T^2+\cdots\right)^{-1}-1~.
\eneq
It is noted that the above equations are valid at all orders, and the
quantities like $T=T^{(1)}+T^{(2)}+\cdots$ are non-perturbative.
At linear order in perturbations, this condition for~$\delta O$
translates into
\beeq
\label{eq:ttobs}
\widetilde{\delta O}^{(1)}(\tilde x)
=\widetilde{\delta O}^{(1)}(x)=\delta O^{(1)}(x)-{d\ln\bar O\over d\eta}~
T^{(1)}~,
\eneq
where we ignore the difference between~$\tilde x$ and~$x$ when computing
the perturbation~$\delta O^{(1)}$ at linear order. Evaluated at the same
coordinate value in two different coordinates (hence the
background correspondence is fixed), this relation is simply
the gauge-transformation relation for~$\delta O^{(1)}$.

While the  relation \eqref{eq:general} is general, in cosmology  we often replace  the background time coordinate by  the observed redshift~$z$, another invariant scalar under the diffeomorphism. Indeed,
the observed redshift provides the  simplest, physically
meaningful way in cosmology to assign a 'time coordinate'  to an event (provided it emits photons). 
Let us repeat the above for the perturbation of the redshift
~$\delta z$. First, we  split the observed redshift into a
background~$\bar z$ and a perturbation~$\dz$ as
\beeq
\label{eq:obsz}
1+z\equiv(1+\bar z)(1+\dz)~,\qquad\qquad 1+\bar z(\eta)\equiv {a(\bar\eta_o)
\over a(\eta)}~,\qquad\qquad \bar \eta_o=\int_0^\infty{dz\over H(z)}~,
\eneq
where $\bar\eta_o$ is the conformal time today in a homogeneous universe,
$H(z)$ is a Hubble parameter, and 
the distortion~$\dz$ encodes the perturbations
due to the peculiar velocity and the potential fluctuation (see 
Appendix~\ref{app:metric}).
We normalize the scale factor~$a$ to unity
at $\bar\eta_o$ so that the redshift parameter~$\bar z$ 
is simply a representation of the time coordinate~$\eta$ 
in a given coordinate system. Though apparent in Eq.~\eqref{eq:obsdef},
we check whether the observed redshift is 
indeed invariant under diffeomorphisms.
We consider the time coordinates $\eta$ and $\tilde\eta=\eta +T$. The background redshift parameters are
related in the two coordinate systems by
\beeq
1+\tilde{\bar z}={a(\bar\eta_o)\over a(\tilde \eta)}={a(\bar\eta_o)\over
a(\eta)}\left(1+\HH~T+\frac12{a''\over a}T^2+\cdots\right)^{-1}=(1+\bar z)
\left(1+\HH~T+\frac12{a''\over a}T^2+\cdots\right)^{-1}~,
\eneq
where $\HH=a'/a$ is the conformal Hubble parameter. This relation dictates 
the transformation properties of the perturbation of the observed redshift
\beeq
\widetilde{\dz}(\tilde x)=(1+\dz(x))
\left(1+\HH~T+\frac12{a''\over a}T^2+\cdots\right)-1~,
\eneq
and at linear order in perturbations  this general relation is indeed
verified by explicit calculations of its gauge-transformation
(see Appendix~\ref{app:metric}):
\beeq
\widetilde{\dz}^{(1)}(\tilde x)=\widetilde{\dz}^{(1)}(x)=\dz^{(1)}(x)
+\HH~T^{(1)}~.
\eneq

As a working example at second order, we consider the luminosity distance~$\dL$.
The background quantity~$\bar D_L(z)$ 
is the luminosity distance in a homogeneous universe:
\beeq
\bar D_L(z)=(1+z)\rbar_z~,\qquad\qquad \rbar_z=\int_0^z{dz\over H(z)}~,
\eneq
where $\rbar_z$ is the comoving distance to the redshift~$z$ in a homogeneous
universe.
Note that in this case the redshift~$z$ is a parameter of the given 
functions~$\rbar$ and~$H$. However, in real observations, 
the luminosity distance~$\dL$ is constructed from observable quantities,
including the observed redshift~$z$ and the observed luminosity of standard candles. We split it into a background~$\bar D_L$
and a perturbation~$\ddL$  using the observed redshift~$z$:
\beeq
\dL(z,\bdi{n})\equiv\bar D_L(z)(1+\ddL)~.
\eneq
The condition that the observed luminosity distance~$\dL$ is an
invariant scalar
implies that the perturbation~$\ddL$ should follow the relation
\beeq
\label{eq:zzobs}
\widetilde{\ddL}(\tilde x)=\ddL(x)~,
\eneq
because the observed redshift~$z$ is also invariant  
under diffeomorphisms,  the correspondence to the background~$\bar D_L(z)$ is 
identical in both coordinate systems.
This relation for the perturbation implies that the perturbation~$\ddL^{(1)}$ 
is gauge-invariant at  linear order in the perturbations,
\beeq
\widetilde{\ddL}^{(1)}(x)=\ddL^{(1)}(x)~,
\eneq
where we again ignore the difference between~$\tilde x$ and~$x$ since this would be  second order. 
Compared to the general relation derived in Eq.~\eqref{eq:ttobs},
this is a special case, which is however generally valid:
the perturbation of an arbitrary (scalar) cosmological observable, expressed in terms of the {\it observed redshift} is gauge invariant at first order.

In summary, the diffeomorphism invariance of observable quantities implies the
gauge-invariance of their perturbation at linear
order.  Beyond  linear order, however, perturbations
are {\it not} gauge-invariant (but still  invariant scalars under
diffeomorphisms).
Given the condition in Eq.~\eqref{eq:zzobs},
the second-order perturbation part gauge-transforms as
\beeq\label{e:rel2}
\widetilde{\delta\mathcal{D}_L^{(2)}}(x)
=\ddL^{(2)}(x)-\xi^{a(1)}\partial_a\ddL^{(1)}~.
\eneq
This relation replaces gauge-invariance for second order perturbations of any variable
which is gauge invariant at first order.  It is the 'sanity check' 
we can use to verify our second order calculations. 
Furthermore, it relates results obtained in different gauges.

\subsection{The Stewart-Walker Lemma}
In this section we connect our discussion to the Stewart-Walker lemma~\cite{Stewart:1974uz}, a well known result from linear cosmological perturbation theory: Consider an observable $O$. Under a linearized coordinate transformation, $\tilde x^a = x^a+\xi^a$, it transforms into
\beeq
\tilde O(x) = O(x) - \LL_\xi O(x) \,,
\eneq
where $\LL_\xi$ denotes the Lie derivative in direction $\xi$.
Splitting $O$ into a background and a perturbation, $O=\bar O + \delta O$,
the first order perturbation $\delta O^{(1)}$ transforms as
\beeq
\widetilde{\delta O}^{(1)}(x)
=\delta O^{(1)}(x) -\LL_{\xi^{(1)}} \bar O(x) \,.
\eneq
The first order gauge transformation is determined by the Lie derivative of 
the background variable in the direction of the (first order) 
displacement $\xi^{(1)}$.
The perturbation variable $\delta O^{(1)}$ is invariant under all gauge 
transformations if and only if $\bar O(x)\equiv 0$ (or constant). 
This lemma can easily be generalized to higher order 
perturbations~\cite{BRMAET97,BRSO99}. 
At order $n$ the perturbation $\delta O^{(n)}(x)$ transforms under $\xi = 
\xi^{(1)}+ \xi^{(2)} + \cdots \xi^{(n)} + \cdots$ as
\bear
\widetilde{\delta O}^{(n)} (x)
&=&\delta O^{(n)}  (x) -\LL_{\xi^{(n)} } \bar O(x)  -\LL_{\xi^{(n-1)} }
\delta O^{(1)}  (x) -\cdots  -\LL_{\xi^{(1)} }\delta O^{(n-1)}(x) \nonumber\\
 && +\frac12\LL _{\xi^{(1)} }\LL_{\xi^{(n-1)} }\bar O(x)
+\frac12\LL_{\xi^{(1)} }\LL_{\xi^{(n-2)} }\delta O^{(1)}(x) + 
\cdots  + \nonumber\\
  &&   +\cdots +\frac{1}{n!}\left(-\LL_ {\xi^{(1)} }\right)^n\bar O \,.  
\enar
For example at second order  this gives
\beeq\label{e:gt2}
\widetilde{\delta O^{(2)}}(x)
=\delta O^{(2)}(x) -\LL_{\xi^{(2)} } \bar O(x)  -\LL_{\xi^{(1)} }\delta 
O^{(1)}(x)   \;+\; \frac12\left(\LL_ {\xi^{(1)} }\right)^2\bar O(x)\,.
\eneq
Hence a second order perturbation variable is gauge invariant only if its background and first order counter parts vanish (or are constant). For a variable which is gauge invariant at first order Eq.~\eqref{e:gt2} implies again \eqref{e:rel2}.

The higher order Steward Walker lemma therefore states:\\
A perturbation variable is gauge invariant at order $n$ if and only if
all its lower order perturbations and its background component vanish (or are constant).
As an example let us again consider the redshift. This is clearly an observable and a diffeomorphism invariant scalar. But as is well known, we show this explicitly also in Appendix~\ref{app:metric}, it is not gauge-invariant, not even at first order as $\bar z\neq 0$. The split of the observed redshift into a background component and a perturbation is arbitrary and not intrinsic.

The situation is somewhat different if we consider the function $\mathcal{D}_L(z, \bdi{n})$.  Strictly speaking this quantity is a bi-scalar, depending on the coordinates $x_o$ of the observer and $x_s$ of the source. Once the observer position is fixed, the fact that the source must lie on the observers background light-cone, together with the observed redshift $z$  and the direction of observation $\bdi{n}$ fully determine the source position. Since $\bar D_L$ does not depend on direction, the higher multipoles of an expansion of $\mathcal{D}_L(z, \bdi{n})$ into spherical harmonics are automatically gauge invariant at 
first order (their background contribution vanishes). But they are not 
gauge-invariant at second order.

The situation is more complicated for the monopole,
\beeq
\langle\delta\mathcal{D}_L(z,\bdi{n})\rangle_\Omega = 
\frac1{4\pi}\int_{\Sb^2}\delta\mathcal{D}_L(z, \bdi{n})d\Omega_{\bdi{n}} \,.
\eneq
Without further care this first order perturbation variable  is not gauge 
invariant as its background contribution  does not vanish. However, it has 
been shown in Ref.~\cite{BIYO16}, that one can choose a physical time 
coordinate at the observer such that
this distance becomes gauge-invariant.

With this remarks the main difference between gauge-invariance and general 
diffeomorphism invariance becomes clear. The former is related to our ability 
to split a variable into a background component and perturbations in a way 
which is independent of the coordinate system. While this is a very useful 
concept 
at the level of first order perturbation theory, it becomes cumbersome and 
not very helpful at higher orders.  For an arbitrary quantity~$O$,
any spatial gradients
\beeq
\nabla_IO \equiv e^a_I\partial_aO~,
\eneq
are pure perturbations and therefore gauge invariant at first order,
providing physical significance to gauge-invariance for linear perturbation 
theory.  
Most common gauge-invariant perturbation variables are actually of this form. 
For the sake of simplicity, we will restrict ourselves to the linear-order
perturbations in constructing the gauge-invariant combinations in this section,
and we will omit the superscript for the perturbation order.
Well known examples are the different gauge invariant combinations for 
the density perturbation,
\bear
D_s &=& \delta +3(1+w)H\chi~,\\
D_g &=& \delta +3(1+w)\phi~, \\
D &=& \delta +3(1+w)\HH (v-\beta)~. \label{e:Dcom}
\enar
Here $\delta$ is the density fluctuation, $v$ is the peculiar velocity potential, $\chi$ and $\phi$ are metric perturbations. The detailed definitions of all variables are given in Appendix~\ref{app:metric}.

Denoting the projection operator into the 3-space normal to the cosmic velocity field by
\beeq
P_a^b = \delta^b_a+u_a u^b~,
\eneq
a short computation gives~\cite{Durrer:1993db}
\bear
D_{,i} &=& P^a_i\rho_{,a}\\
(D_s)_{,ij} +3(1+w)\Psi_{,ij}&=& P^b_jP^a_i\rho_{;ab}  \\
(D_g)_{,ij} +3(1+w)(\Psi+\Phi)_{,ij}&=& P^b_jP^a_i\rho_{;ab} \,,
\enar
where  $\Psi$ and $\Phi$ are the Bardeen potentials. 
Hence $D$ and $D_s+3(1+w)\Psi$ are potentials for the first and second gradient of the density projected onto the 3-space normal to $u^a$. 
Similarly, the sum $\Phi+\Psi$ is the potential for the 
Weyl curvature and $\Phi-\Psi$ is a potential for the 
fluid anisotropic stress which both vanish in a FLRW spacetime.

As such, gauge invariance can play an important role for first order 
perturbation 
theory but much less so for second and higher order. For this reason, 
at higher orders we advocate to use diffeomorphism invariance which 
is of course maintained order by order.

\section{Application to the light-cone average}
\label{sec:obsavg}
As an application of our findings of the  previous section, we derive 
the expression for the observed light-cone average and identify a few 
relativistic corrections that are absent in previous work. We 
show that our expression is also an invariant scalar.
'Light cone average' means the average over directions for objects on the light-cone at fixed observed redshift, i.e., it is an angular average over the two dimensional hypersurface of constant redshift on the observer's 
background light-cone. 
This approximates the average done in actual observations.
Here we relate it to the ensemble average  because this is the quantity which we can calculate. As an example, we use again the luminosity distance but for any other quantity defined on the background light-cone the procedure is equivalent.

\subsection{Neglecting volume and source fluctuations}
While observers perform the light-cone average
on the uniform surface of a sphere with radius $\rbar_z$,
the true  radius depends on directions leading to an inhomogeneous surface, 
and this surface depends on the choice of coordinate systems (see
 Appendix~\ref{app:metric}). The observational
angular average needs to be expressed in terms of an average of objects over
this inhomogeneous surface.

In a given coordinate system, we parametrize the position of the objects
in terms of their observed redshift~$z$ as
\beeq
\label{eq:src}
x^a_s=[\eta_z+\DT,~\bar x^i_z+\DX^i]~,\qquad \qquad 1+z\equiv
{a(\bar\eta_o)\over a(\eta_z)}~,\qquad\qquad \bar x_z^i=\rbar_z n^i~.
\eneq
Here $\eta_z$ is defined by the middle identity and $\bar r_z$ is the comoving 
distance out to redshift $z$ in the background.
The spatial displacement $\DX^i$ 
of the source position with respect to the position
inferred based on the observed redshift~$z$ and the observed direction~${\bdi n}$ are
 often expressed in  spherical coordinates as ($\drr,\dtt,\dpp$).  Since the background
correspondence is fixed by using the observed redshift (invariant scalar),
the displacements  in two different coordinate systems are 
non-perturbatively related by
\beeq
\widetilde{\DT}(\tilde x)=\DT(x)+T(x)~,\qquad\qquad 
\widetilde{\DX}^i(\tilde x)=\DX^i(x)+L^i(x)~.
\eneq
At linear order  these relations are the
gauge-transformation equations and indeed 
when the displacements are expressed in terms of metric 
perturbations, they do gauge-transform exactly  as expected
 (see Appendix~\ref{app:metric})
\beeq
\widetilde{\DT}^{(1)}(x)=\DT^{(1)}(x)+T^{(1)}(x)~,\qquad\qquad 
\widetilde{\DX}^{i(1)}(x)=\DX^{i(1)}(x)+L^{i(1)}(x)~.
\eneq

Suppose we measure the luminosity distances from infinitely many supernovae
in the sky and they are all at the same observed redshift. In a first step, we assume 
that these objects are uniformly distributed on observer directions $\bdi n$, such that
our observational
angular average of the luminosity distance at the same observed redshift
is equally weighted for each object (in reality,  objects are weighted differently due to  volume distortions and due to source fluctuations,
which we discuss in Sec.~\ref{sec:relation}). In this case, we need
to express the real source position in Eq.~\eqref{eq:src} around the uniform
sphere and average the luminosity distance over the uniform sphere,
accounting for its distortion. This procedure amounts to averaging the
luminosity distance over the inhomogeneous sphere. Up to the second
order in perturbations, the luminosity distance is written as
\beeq
\label{eq:inv}
\dL(z,\bdi{n};x^a_s)=\bar D_L(z)
\bigg(1+\ddL^{(1,2)}(\bar x^a_z)+\DX^{a(1)}\partial_a\ddL^{(1)}
(\bar x_z^a)+\OO(3)\bigg)~,
\eneq
and when averaged over the uniform sphere~$\bar x_z^i$ (or a spatial 
coordinate) we can relate it to the ensemble average in the given coordinate,
\beeq
\label{eq:angavg}
\AVE{\dL(z,\bdi{n})}_\Omega=\bar D_L(z)\bigg(1+
\underbrace{\AVE{\ddL^{(1)}(\bar x^a_z)}}_{=0}+\AVE{\ddL^{(2)}(\bar x^a_z)}
+\AVE{\Delta x^{a(1)}\partial_a\ddL^{(1)}(\bar x_z^a)}+\OO(3)\bigg)~.
\eneq

To ensure that this expression is correct, we check diffeomorphism invariance of
the observational angular average  to the second order in perturbations, as this
observational procedure is again independent of the coordinates used to compute it. In the previous
section we derived that the second-order perturbation $\ddL^{(2)}$ of the luminosity distance transforms as
\beeq
x^a\ra \tilde x^a = x^a+\xi^a, \qquad  \ddL^{(2)}(x) \ra
\widetilde{\ddL}^{(2)}(x)=\ddL^{(2)}(x)-\xi^{a(1)}\partial_a\ddL^{(1)}~.
\eneq
The correction terms in Eq.~\eqref{eq:angavg}
due to the deviation of the source position gauge-transform as
\beeq
\widetilde{\DX}{}^{a(1)}\partial_a\widetilde{\ddL}{}^{(1)}
=\left(\DX^a+\xi^a\right)^{(1)}\partial_a\ddL^{(1)}~,
\eneq
where $\ddL$ is gauge-invariant at the linear order. Adding these two 
contributions, we readily derive that 
the sum of the two terms in our angular average is indeed
invariant under  diffeomorphisms.

\subsection{Including volume and source fluctuations}
\label{sec:relation}

As briefly discussed in the previous section, objects on an inhomogeneous
sphere are in fact {\it not} equally weighted. In the observational light-cone
average,
equal weights are given to  objects within
the same background volume element $d\bar V$, determined by the
observed redshift~$z$ and the observed direction~$\bdi{n}=(\ttt,\pp)$:
\beeq
d\bar V(z)\equiv{\rbar^2(z)\over H(z)(1+z)^3}~dz~d\Omega~,\qquad\qquad
d\Omega=\sin\ttt~d\ttt d\pp~.
\eneq
In other words, objects within the same observed redshift bin~$dz$
are equally counted over the observed angle. Due to the inhomogeneities
in our universe, the physical volume $dV_\up{phy}$ that appears as 
the observed volume~$d\bar V$ is different at each point, and hence
each point obtains different weights. This physical volume in the source rest frame
is given by~\cite{WEINB72}
\beeq
\label{eq:volP}
dV_\up{phy}=\sqrt{-g}~\varepsilon_{dabc}u_s^d{\partial x_s^a\over\partial z}
{\partial x_s^b\over\partial \ttt}{\partial x_s^c\over\partial \pp}
dz~d\ttt~d\pp\equiv (1+\delta V)d\bar V(z)~,
\eneq
where we define the volume distortion~$\delta V$, $\varepsilon_{abcd}$ is the 
Levi-Civita symbol ($\varepsilon_{0123}=1$), the metric determinant
is~$g$, and the source velocity is~$u^a_s$. As is apparent from the definition,
the volume distortion is another invariant scalar at all orders,
\beeq
\widetilde{\delta V}(\tilde x)=\delta V(x)~,
\eneq
and it is gauge-invariant at linear order $\widetilde{\delta V}{}^{(1)}(x)
=\delta V^{(1)}(x)$.
This covariant expression for the physical volume 
 is essential in deriving the correct relativistic
formula for galaxy clustering \cite{YOFIZA09,YOO10,BODU11,JESCHI12,YOO14a}.
The dominant contributions to the volume distortion~$\delta V$ are 
redshift-space distortions and gravitational lensing, but there exist 
other relativistic effects~\cite{BODU11}.
Accounting for the different weight 
due to the volume distortion, the observational light-cone
average can be expressed in terms of the observational angular average as
\beeq
\langle\ddL\rangle_{\rm obs^{(V)}}(z)={1\over\int dV_\up{phy}}\int dV_\up{phy}~\ddL(z,\bdi{n} )
={\left\langle\ddL(z,\bdi{n})(1+\delta V)\right\rangle_\Omega\over
\left\langle1+\delta V\right\rangle_\Omega}~,
\eneq
where we use the fact that the observational average is performed over
the same observed redshift bin~$dz$. Note that we used the superscript~$V$
to indicate that we account for the volume weight.

One last ingredient for the complete description 
of~$\langle\ddL\rangle_{\rm obs}$ 
is to consider the clustering of
sources~\cite{YOO09}. Apart from the volume distortion~$\delta V$,
fluctuations in the source number density also affect our observable.
In our example, supernova hosts are remote galaxies, and these galaxies
are clustered. Furthermore, observational selection such as the magnitude
threshold can further bias the observed sample, and all these effects associated
with the source galaxies are collectively called the source effect.
Similar arguments for the source effect in  light-cone averaging
are  discussed in \cite{KAHU15a,FLCLMA16}. 
Together with the volume distortion,
the observational light-cone average indeed gives equal weight to the observed
counts $dN_g^\up{obs}$ of  sources (or galaxies) in the observed 
volume~$d\bar V$. This number of  observed galaxies is used to define the
observed galaxy number density~$n_g^\up{obs}$ and is related to the 
physical galaxy number density~$n_g^\up{phy}$ and the volume 
distortion~$\delta V$ as
\beeq
dN_g^\up{obs}(z,\bdi{n})=n_g^\up{obs}(z,\bdi{n})d\bar V=n_g^\up{phy}dV_\up{phy}
=\bar n_g(z)(1+\delta_g)(1+\delta V)~d\bar V~,
\eneq
where we define the source fluctuation by splitting the physical galaxy
number density $n_g^\up{phy}$ into the background $\bar n_g(z)$
at the observed redshift and the remaining fluctuation~$\delta_g$ around it.
For the same reason, the intrinsic source fluctuation is an invariant scalar
at all orders,
\beeq
\widetilde{\delta}_g(\tilde x)=\delta_g(x)~,
\eneq
and it is gauge-invariant at linear order in perturbations
$\widetilde{\delta}{}^{(1)}_g(x)=\delta_g^{(1)}(x)$.
The intrinsic source fluctuation
is the density fluctuation in comoving gauge~$D^{(1)}$ in 
Eq.~\eqref{e:Dcom} up to some bias factor and the perturbation~$\dz$
to compensate the difference between the observed redshift and the
proper time of the source galaxies \cite{YOO14a}.
Therefore, the observational light-cone average~$\langle\ddL\rangle_{\rm obs}$ 
remains unaffected under a diffeomorphism when
adding the additional corrections due to the volume
distortion~$\delta V$ and the source fluctuation~$\delta_g$.

Including both, the volume and source fluctuations, 
the observational light-cone average is given by
\beeq
\label{eq:final}
\langle\ddL\rangle_{\rm obs}={1\over \int dN_g^\up{obs}}\int dN_g^\up{obs}~\ddL(z,\bdi{n})
={\left\langle\ddL(z,\bdi{n})(1+\delta_g)(1+\delta V)\right\rangle_\Omega\over
\left\langle(1+\delta_g)(1+\delta V)\right\rangle_\Omega}~,
\eneq
where the denominator  in the first equality 
is just the total number of galaxies within the survey volume $(\Omega,~dz)$.
This expression is equivalent to that derived
in \cite{FLCLMA16}, while only the dominant terms are considered
in \cite{KAHU15a} where the relativistic contributions to~$\delta_g$ 
and~$\delta V$ are ignored. Finally, we
relate the observed angular average to the ensemble average by perturbatively
expanding the contributions. To  second order in perturbations,
the observational light-cone average is 
\bear
\Big\langle\langle\ddL\rangle_{\rm obs}\Big\rangle
&=&\Big\langle\left\langle\ddL\right\rangle_\Omega+\left\langle\ddL(\delta_g+\delta V)
\right\rangle_\Omega\Big\rangle+\OO(3)  \nonumber\\
&&\hspace{-2cm}= ~ \left\langle\ddL^{(2)}(\bar x^a_z)\right\rangle +
\left\langle\Delta x^{a(1)}\partial_a\ddL^{(1)}(\bar x_z^a)\right\rangle+
\left\langle\ddL^{(1)}\left(\delta^{(1)}_g+\delta V^{(1)}\right) 
\right\rangle+\OO(3)~,
\enar
where we use the relation of the observational
angular average to the ensemble average
in Eq.~\eqref{eq:angavg}. 

Given the complete description of the observational light-cone average,
we also clarify its relation to the directional average in~\cite{BOCLET15b},
where the observational angular average is related to the ensemble average.
In this discussion, volume distortion~$\delta V$ and 
 source fluctuations~$\delta_g$ have been ignored,
focusing on the angular average in Eq.~\eqref{eq:angavg}.
The difference between the ensemble average and the angular average
is exactly due to the deviation of the source position from the uniform
sphere:
\bear
\AVE{\ddL}_\Omega-\AVE{\ddL} &=&
\left\langle\DX^{a(1)}\partial_a\ddL^{(1)}(\bar x_z^a)\right\rangle
+\OO(2) \\
\label{eq:direction}
&=&\left\langle\Delta\eta^{(1)}~\ddL^{\prime(1)}\right\rangle
+\left\langle\drr^{(1)}{\partial\over\partial\rbar}\ddL^{(1)}\right\rangle
+\left\langle\left(\dtt^{(1)}{\partial\over\partial\ttt}
+\dpp^{(1)}{\partial\over\partial\pp}\right)\ddL^{(1)}
\right\rangle+\OO(3)~.\nonumber
\enar
The angular distortions in the
round bracket can be re-arranged as a total  derivative and
the divergence of the angular distortions as
\beeq
\left(\dtt^{(1)}{\partial\over\partial\ttt}
+\dpp^{(1)}{\partial\over\partial\pp}\right)\ddL^{(1)}=\hat\nabla\cdot
\left[(\dtt,~\dpp)^{(1)}\ddL^{(1)}\right]-\ddL^{(1)}\hat\nabla\cdot(\dtt,
~\dpp)^{(1)}~.
\eneq
The integral over the  total divergence vanishes and the second term is the gravitational lensing convergence
\beeq
-2\kappa\equiv \hat\nabla\cdot(\dtt,~\dpp)^{(1)}=
\left(\cot\ttt+{\partial\over\partial\ttt}
\right)\dtt^{(1)}+{\partial\over\partial\pp}\dpp^{(1)}~.
\eneq
Averaging over angles, 
we find that the last term in Eq.~\eqref{eq:direction}
is indeed the expression found in
Ref.~\cite{BOCLET15b}, in which only the dominant contributions to the average
were considered. In addition to this lensing correction to the angular
average, there exist
the radial distortion~$\drr$ and the displacement~$\Delta\eta$ in the time
coordinate due to the mismatch between the background source position
inferred from its redshift and the real source position. 
However, these contributions are in general
smaller than the gravitational lensing convergence as noted in 
\cite{BOCLET15b}.

\section{Discussion}
\label{sec:discuss}
Cosmological information is carried to us by electromagnetic waves (or gravitational
waves) from distant sources, and this information is measured by an observer
in her rest frame. The observer four velocity sets the time direction,
and an orthonormal triad establishes the spatial directions in the
observer's rest frame.  This frame is unique up to a rotation $R\in SO(3)$ of the triad $\bdi{e}_i$. This unique setting specifies the frame in which
cosmological information is decoded and stored. 
For example,  photon frequencies are measured
in the observer's rest frame, and photon propagation directions are
measured against the spatial triad  in the observer's rest frame.
Therefore, cosmological observables that encode physical properties
of the sources, the geometry of spacetime and the nature of gravity are 
expressed in terms of scalars (under diffeomorphisms) 
defined along the world line of the observer.

In contrast to the observer's rest frame,
 theoretical predictions for  cosmological observables are described 
in an arbitrary coordinate system, and general covariance
ensures that any choice of coordinates can be
used to describe cosmological observables. 
Changes in the coordinates have {\it no} impact on the unique frame of
the observer for cosmological observations. Therefore, 
 cosmological observables  have
 identical values, regardless of our choice of coordinates.
At linear order, this redundancy is often phrased
as  gauge freedom and is removed by demanding that
theoretical predictions for cosmological observables be gauge-invariant.
However, this statement is valid only for linear  perturbations,
because the non-perturbative condition for cosmological observables is that
they remain invariant under diffeomorphisms.
Consequently, theoretical predictions beyond the linear order do
{\it gauge-transform} in a unique way (i.e., they are in general not gauge-invariant),
and their transformation property, given in Eq.~\eqref{e:rel2} for  second order perturbations, can be used to check the validity
of  theoretical predictions beyond  linear
order in perturbations.
Very few cosmological observables actually do have vanishing first order contribution and therefore are gauge invariant at second order. An example is the rotation angle in the lens map for the case of vanishing first order vector and tensor perturbations~\cite{Marozzi:2016und}.

We have applied our findings to the observational light-cone average of the
luminosity distance. As one of the cosmological observables, the observational
light-cone average is an invariant scalar. 
In previous work \cite{BOCLET15b,FLCLMA16,KAPE16}
on the observational light-cone
average, even though  correctly accounting for the dominant contributions to the second order
in perturbations,   diffeomorphism-invariance is violated.
Improving upon this work, we have provided a complete description of
the observational light-cone average, including the relativistic corrections
and checking the invariance of the observational light-cone
average under  diffeomorphisms in FLRW coordinates.

Similar considerations have been performed almost two decades ago \cite{BRSO99}
(see also \cite{BRMAET97}).
Noting that while perturbations are in general gauge-dependent,
the perturbation part of observable quantities such as the cosmic
microwave background
anisotropies is gauge-independent, they have adopted the view that
``an observable quantity in general relativity is simply represented by
a scalar field on spacetime,'' and proceeded to find that once the
background is gauge-independently defined, the perturbation part is 
gauge-invariant at  first order, but transforms at  higher orders
in perturbations. 
In our language, they assumed that cosmological observables and
the background quantities of such observables 
are invariant scalars under diffeomorphisms.
Here we explicitly prove that cosmological observables measured by the observer in her rest frame
are indeed scalars under diffeomorphisms when the background quantities are expressed in terms of the observed
redshift.
Given this identification, the work
by \cite{BRSO99} naturally leads to the conclusions of the present paper.

\acknowledgments
We acknowledge useful discussions 
with Camille Bonvin, Chris Clarkson, Pierre Fleury,
Jinn-Ouk Gong, and
Roy Maartens. We thank Ermis Mitsou for clarifying the global and the local
symmetries involved in the calculations and Marco Bruni for pointing us to
his early work.
We acknowledge support by the Swiss National Science Foundation.
J.Y. is also supported by
a Consolidator Grant of the European Research Council (ERC-2015-CoG grant
680886).

\appendix
\section{Metric convention and gauge-transformation}
\label{app:metric}
Here we briefly summarize our metric conventions and present equations
for the fluctuations of the observed redshift and the source position
(see \cite{YOO14a} for detailed derivations). The metric of an inhomogeneous universe close to a FLRW universe is modelled by four scalar perturbations 
$(\alpha,\beta,\varphi,\gamma)$ 
\beeq
ds^2=-a^2(1+2\alpha)d\eta^2-2a^2\beta_{,i}d\eta dx^i+a^2\left[(1+2\varphi)
\delta_{ij}+2\gamma_{,ij}\right]dx^idx^j~,
\eneq
where $a(\eta)$ is the scale factor, $\delta_{ij}$ is the Kronecker delta,
and we ignore vector and tensor perturbations as well as spatial curvature for simplicity.
This metric representation is 
non-perturbative and fully general, e.g., $\alpha=\alpha^{(1)}+\alpha^{(2)}
+\cdots$. However, in this Appendix we will only deal with the {\it
linear}
perturbations. Hence we will omit the superscript in the Appendix.
Under the general coordinate transformation in Eq.~\eqref{eq:ct}
these linear scalar perturbations transform as
\beeq
\tilde\alpha=\alpha-{1\over a}\left(aT\right)'~,\qquad 
\tilde\beta=\beta-T+L'~,\qquad
\tilde\varphi=\varphi-\HH T~,\qquad\tilde\gamma=\gamma-L~,
\eneq
where the prime is the derivative with respect to the conformal time~$\eta$
and the scalar ~$L$ is  defined by $L^i\equiv \delta^{ij}\partial_j L$. For a later convenience, we introduce
$$\chi\equiv a(\beta+\gamma')$$ and  the velocity potential~$v$ such that the observer four velocity~$u^a$ is given by
$$u_i\equiv-v_{,i}~.$$ These transform  as 
\beeq
\tilde \chi=\chi-aT~,\qquad\qquad
\tilde v=v-T~.
\eneq
Given our metric convention and the gauge transformation properties,
we can derive the expressions for the linear fluctuations of the redshift and the coordinates $\Delta x^a$ introduced in Eq.~\eqref{eq:src} by solving the photon geodesic equation. The perturbation of
 the observed redshift defined as in Eq.~\eqref{eq:obsz} is
\beeq
\dz=-H\chi+(H_o\chi_o+\HH_o\dT_o)+\left[V-\ax\right]_s
-\left[V-\ax\right]_o-\int_0^{\rbar_z}
d\rbar~ (\ax-\px)'~,\qquad\widetilde{\dz}=\dz+\HH T~,
\eneq
where the subscripts~$o$ and~$s$ represent the observer and the source
positions, $\delta\eta_o$ is the coordinate lapse of the observer time
coordinate, 
\beeq
\dT_0=-\int_0^{\bar\eta_o}d\eta~a~\alpha~,\qquad\qquad
\widetilde{\dT}_o=\dT_o+T_o~,
\eneq
and the equation is re-arranged to isolate the gauge-dependence by
using the gauge-invariant quantities:
\beeq
\ax\equiv \alpha-\frac1a\chi'\equiv\Psi ~,\qquad \px\equiv\varphi-H\chi
\equiv -\Phi ~,\qquad
V\equiv-{\partial\over\partial r}v_\chi~,\qquad v_\chi\equiv v-\frac1a\chi~.
\eneq
Here $\Psi$ and $\Phi$ are the Bardeen potentials.
Compared to the inferred position of the source at the observed redshift,
the fluctuation of conformal time $\Delta\eta$ is related to the 
 redshift perturbation as
\beeq
\Delta\eta={\dz\over\HH}~,\qquad\qquad\widetilde{\Delta\eta}=\Delta\eta
+T~.
\eneq
The fluctuation of the source position  $(\drr,\dtt,\dpp)$ in
Eq.~\eqref{eq:src} can be expressed as

\bear
\drr&=&-{\partial\gamma\over\partial\rbar}
+n_i\left(\delta x^i+\gamma^{,i}\right)_o+
\left(\chi_o+\dT_o\right)-\frac1\HH\left(\dz+H\chi\right)
+\int_0^{\rbar_z}d\rbar\left(\ax-\px\right)~,\\
\rbar_z\dtt&=&
-{1\over\rbar_z}{\partial\gamma\over\partial\ttt}
+(e_\theta)_i\left(\delta x^i+\gamma^{,i}\right)_o
-\int_0^{\rbar_z}d\rbar\left({\rbar_z-\rbar\over\rbar}\right){\partial
\over\partial\ttt}\left(\ax-\px\right)~,
\enar
where $e_\theta$ is the unit vector in direction of $\theta$ and $\delta x^i_o$ is the spatial coordinate lapse of the observer,
\beeq
\delta x^i_o=\int_0^{\bar\eta_o}d\eta~a~u^i~,
\qquad\qquad \widetilde{\delta x}^i_o=\delta x^i_o+L^i_o~.
\eneq
A similar equation is obtained for the azimuthal fluctuation~$\rbar_z\sin\ttt~\dpp$, see e.g.~\cite{BODU11}.
As is easily checked, the terms in  round brackets are gauge-invariant combinations, and
consequently all the perturbations transform as expected,
\beeq
\widetilde{\drr}=\drr+{\partial L\over\partial\rbar}~, \qquad\qquad
\rbar_z \widetilde{\dtt}
=\rbar_z\dtt+{1\over\rbar_z}{\partial L\over\partial\ttt}~.
\eneq
Finally, the volume distortion defined in \eqref{eq:volP} can be derived as
\beeq
\delta V=
3\left(\dz+H\chi\right)+\ax+2~\px+{2\over\rbar_z}\left(\drr+{\partial\gamma
\over\partial \rbar}\right)
-2\left(\kappa-{1\over2\rbar_z^2}\hat\nabla^2\gamma\right)
-H{\partial\over\partial\zz}\left({\dz+H\chi\over\HH}\right)+V~,
\eneq
which is indeed gauge-invariant. To see this one  has to verify that the convergence $\kappa$ gauge transforms as 
$$\tilde\kappa=\kappa -\frac{\hat\nabla^2 L}{2r_z^2}\,.$$
More details are found e.g. in~\cite{YOO14a,BODU11}.

\bibliographystyle{JHEP}
\bibliography{reference_v6}

\end{document}